\documentclass[conference]{IEEEtran}
\IEEEoverridecommandlockouts
\usepackage{cite}
\usepackage{amsmath,amssymb,amsfonts}
\usepackage{graphicx}
\usepackage{textcomp}
\usepackage{xcolor}
\usepackage{url}
\usepackage{booktabs}
\usepackage{multirow}
\usepackage{array}
\usepackage{pifont}
\definecolor{prismblue}{RGB}{0,70,160}
\usepackage[
    colorlinks=true,
    linkcolor=prismblue,
    citecolor=prismblue,
    urlcolor=prismblue
]{hyperref}

\newcommand{\cmark}{\ding{51}}
\newcommand{\xmark}{\ding{55}}
\newcommand{\best}[1]{\textbf{#1}}
\def\BibTeX{{\rm B\kern-.05em{\sc i\kern-.025em b}\kern-.08em
    T\kern-.1667em\lower.7ex\hbox{E}\kern-.125emX}}
\begin{document}

\title{Transition-Aware Backend Dispatch for Edge LLM Inference}

\author{
Alaaddin Goktug Ayar, 
Martin Margala\\

{School of Computing and Informatics, University of Louisiana at Lafayette, Lafayette, LA, USA}\\

{\small alaaddin.ayar1@louisiana.edu, martin.margala@louisiana.edu}
}

\maketitle

\begin{abstract}
Efficient large language model (LLM) inference on edge platforms is limited not only by model size, but also by shape-dependent performance differences across execution backends. Static backend assignment cannot exploit this variation, while independent per-operator selection can introduce costly device and framework switches. This paper presents a transition-aware backend dispatch approach for edge transformer inference. The approach combines current operator features with the previously selected backend to preserve beneficial shape-specific choices while avoiding unnecessary transitions. Ordered traces are collected from full-model inference runs of seven transformer models, and four common operator classes are benchmarked across PyTorch eager CPU, PyTorch eager CUDA, and ONNX Runtime CPU on an NVIDIA Jetson platform. The dispatch policies are evaluated through measurement-backed trace replay using observed operator costs and transition costs measured from actual backend switches. Supported operators are selected dynamically, while operators outside the dispatch scope retain a static assignment. Across 9,584 ordered operator instances and 278 exact shape groups, transition-aware dispatch reduces replayed latency, energy, and energy-delay product relative to the best static policy by 17.4\%, 14.4\%, and 28.5\% on average, respectively. It also reduces switching relative to operator-local selection. Leave-one-model-out evaluation improves all three objectives for six of seven held-out models and improves energy for all seven. These results demonstrate that incorporating operator shape and backend-transition context can improve selective backend dispatch for edge transformer workloads.
\end{abstract}

\section{Introduction}

Large language models (LLMs) are becoming an important component of interactive and agentic artificial intelligence systems. Although the largest models continue to require datacenter-scale infrastructure, smaller transformer models increasingly enable practical inference on edge platforms. Local execution can reduce network dependence and cloud-serving costs, improve response latency, and preserve privacy by keeping user data on the device. These benefits make efficient transformer inference important for embedded assistants, robotics applications, and real-time decision systems~\cite{Vaswani2017Attention,Touvron2023LLaMA,Kwon2023vLLM,Song2024PowerInfer,Xue2024PowerInfer2, word2hypervec}.

Edge deployment presents different optimization constraints from cloud inference. Edge devices operate under tighter power, memory, and thermal limits while exposing heterogeneous resources such as multicore CPUs, embedded GPUs, and runtime-specific backends~\cite{Lane2016DeepX,Huynh2017DeepMon,Kang2017Neurosurgeon,Dagli2022AxoNN}. A common deployment strategy assigns a single backend to the model or to large portions of its computation graph. Although simple, static assignment cannot exploit performance differences across operator types, tensor shapes, inference phases, and optimization objectives~\cite{Chen2018TVM,Zheng2020Ansor,Shao2022MetaSchedule}.

Transformer inference repeatedly invokes dense projections, multilayer perceptron (MLP) activations, and normalization operations. The relative performance of these operators can vary across tensor shapes and backend implementations. A backend that minimizes latency for one shape may be inefficient for another, and the backend that minimizes latency may differ from the one that minimizes energy or energy-delay product (EDP). Static assignment can therefore miss useful performance and energy-efficiency opportunities on heterogeneous edge platforms~\cite{Dagli2022AxoNN,Seo2021SLOAware,Kang2021LaLaRAND}.

Operator-local selection provides a more flexible alternative by assigning each supported operator to its lowest-cost backend. However, independent decisions can produce frequent backend, device, or framework switches over an ordered operator trace. These transitions can introduce synchronization, data movement, and framework-conversion costs that offset the benefit of selecting a faster backend for an individual operator~\cite{Kang2017Neurosurgeon,Dagli2022AxoNN,Kang2021LaLaRAND,Jeong2022Band}. Consequently, a dispatch policy should consider both the current operator cost and the backend selected for the preceding operator.

This paper presents a selective transition-aware backend dispatch approach for edge transformer inference. The selector uses current operator features together with the previous backend state to avoid transitions whose cost exceeds their local benefit. Three eager-mode execution backends are considered: PyTorch CPU, PyTorch CUDA, and ONNX Runtime CPU. Supported operators are assigned dynamically, while operators outside the dispatch scope retain a static backend assignment.

The approach is evaluated through measurement-backed replay of ordered traces collected from full-model inference runs. Exact operator shapes and backend transitions are measured on an NVIDIA Jetson platform, and the resulting costs are replayed in the original model-derived order. This methodology evaluates the performance potential of selective eager-mode dispatch without requiring an integrated mixed-framework runtime.

The evaluation covers seven transformer model traces, 9,584 ordered instances of four operator classes, and 278 exact shape groups. Transition-aware dispatch reduces replayed latency, energy, and EDP relative to the best static backend policy by 17.4\%, 14.4\%, and 28.5\% on average, respectively. It also reduces switching relative to operator-local selection. Leave-one-model-out evaluation improves all three objectives for six of seven held-out models and improves energy for all seven.

The main contributions are as follows:

\begin{itemize}
    \item A full-model-derived operator-trace benchmark covering seven transformer models, 9,584 ordered operator instances, and 278 exact shape groups.

    \item Exact-shape measurements across PyTorch CPU, PyTorch CUDA, and ONNX Runtime CPU, including latency, power, energy, EDP, correctness validation, and measured backend-transition costs.

    \item Operator-local and transition-aware selectors for latency, energy, and EDP, with previous-backend state incorporated into transition-aware decisions.

    \item A measurement-backed trace-replay evaluation that quantifies replay-cost improvements, switch reductions, and leave-one-model-out transfer.

    \item An open-source implementation and reproducibility workflow that supports additional models, operators, and execution backends~\cite{PowerAwareEdgeInferencePublic2026}.
\end{itemize}

\section{Background and Motivation}

This section motivates transition-aware backend dispatch by examining the constraints of edge transformer inference, the opportunities and coordination costs introduced by heterogeneous execution, and the shape-dependent behavior of common transformer operators. Together, these factors motivate a selective dispatch policy that considers both individual operator costs and transitions between consecutive backend assignments.

\subsection{Edge LLM Inference}

LLM inference systems have traditionally targeted server-class GPUs and cloud-serving infrastructure, with substantial attention given to batching, memory management, model offloading, and serving throughput~\cite{Yu2022Orca,Aminabadi2022DeepSpeedInference,Sheng2023FlexGen,Kwon2023vLLM}. Edge deployment presents a different optimization problem because devices operate under tighter power, memory, and thermal limits while still requiring responsive execution for user-facing applications~\cite{Lane2016DeepX,Dagli2022AxoNN}.

The increasing availability of smaller transformer models makes local execution practical for more workloads. Tasks that do not require large cloud-hosted models can benefit from on-device execution when model and runtime behavior are optimized for the target platform. Recent work has studied efficient LLM execution under constrained-memory, consumer-GPU, mobile, and collaborative edge settings~\cite{Song2024PowerInfer,Xue2024PowerInfer2,Alizadeh2024LLMInAFlash,Chen2024EdgeShard, goktug_llm}. This motivates studying NVIDIA Jetson, where operator shape, processing device, and software backend can strongly influence latency and energy behavior.

\subsection{Heterogeneous Execution on Edge Platforms}

Edge platforms expose heterogeneous execution resources, including general-purpose CPUs, embedded GPUs, accelerators, and runtime-specific backends. Prior work on mobile and embedded inference has shown that processor-aware execution can improve latency, energy efficiency, or service-level behavior compared with assigning an entire workload to a single resource~\cite{Lane2016DeepX,Huynh2017DeepMon,Kang2017Neurosurgeon,Seo2021SLOAware,Dagli2022AxoNN,Kang2021LaLaRAND,Jeong2022Band}. These studies establish backend assignment as an optimization decision rather than a fixed deployment choice.

Heterogeneous execution also introduces coordination costs. Moving between devices or runtime frameworks may require synchronization, data movement, tensor conversion, and changes in execution context. Consequently, selecting a backend using only the isolated cost of the current operator may produce locally favorable decisions whose switching overhead reduces the overall benefit~\cite{Dagli2022AxoNN,Kang2021LaLaRAND,Jeong2022Band}. A transition-aware policy must therefore account for both operator cost and the backend used by the preceding operator. Prior heterogeneous inference systems primarily schedule complete models, layers, or graph partitions across processing resources~\cite{Dagli2022AxoNN,Kang2021LaLaRAND,Jeong2022Band}. The present study instead focuses on repeated exact-shape transformer operators and considers transitions across both execution devices and eager-mode runtime frameworks.

\subsection{Shape-Dependent Operator Behavior}

Transformer inference repeatedly invokes operations such as dense projections, feed-forward activations, normalization, and attention-related computation~\cite{Vaswani2017Attention,Devlin2019BERT,Touvron2023LLaMA}. Their performance depends on tensor dimensions, sequence length, hidden size, data type, and backend implementation. A backend that performs well for a large matrix operation may be inefficient for a smaller activation or normalization operation, particularly on resource-constrained edge devices.

Compiler and autotuning frameworks have long recognized that tensor performance is shape- and hardware-dependent. Existing approaches perform graph lowering, schedule search, learned cost modeling, graph substitution, and automatic tensor optimization~\cite{Chen2018TVM,Chen2018AutoTVM,Zheng2020Ansor,Shao2022MetaSchedule,Feng2023TensorIR,Jia2019TASO,Ma2020Rammer}. Unlike systems that generate new schedules or kernels, the present study selects among existing eager-mode backend implementations using exact-shape costs measured on the target device. This formulation is intended for deployment settings built from established frameworks and runtime libraries.

\subsection{Need for Transition-Aware Backend Dispatch}

Three dispatch strategies motivate the proposed design. Static assignment retains one backend and avoids switching, but cannot exploit shape-dependent backend preferences. Operator-local selection chooses the lowest-cost backend independently for each supported operator, but can introduce frequent device and framework transitions. Transition-aware dispatch adds the previous backend state to each supported-operator decision, allowing the policy to retain beneficial shape-specific assignments while avoiding transitions whose measured cost exceeds their local benefit.

The proposed policy is evaluated through measurement-backed replay of model-derived operator traces. This setting isolates the effect of backend choice and transition behavior while preserving the original operator order.

\begin{figure*}[t]
    \centering
    \includegraphics[width=\textwidth]{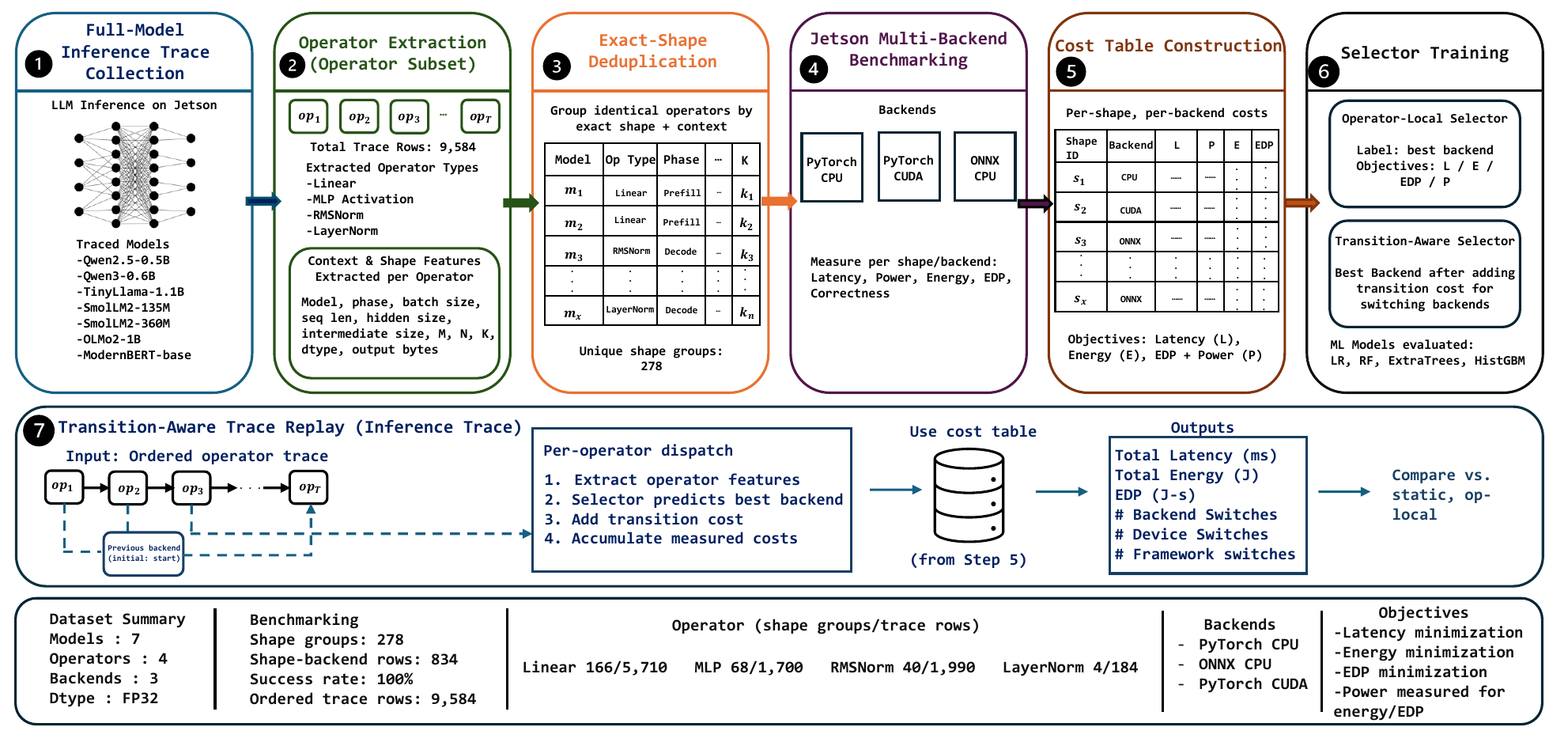}
    \caption{Overview of the proposed transition-aware backend dispatch framework, from full-model inference trace collection to exact-shape benchmarking, selector training, and trace replay evaluation.}
    \label{fig:1stfigure}
    
\end{figure*}

\section{Methodology}

\begin{figure}[t]
    \centering
    \includegraphics[width=\columnwidth]{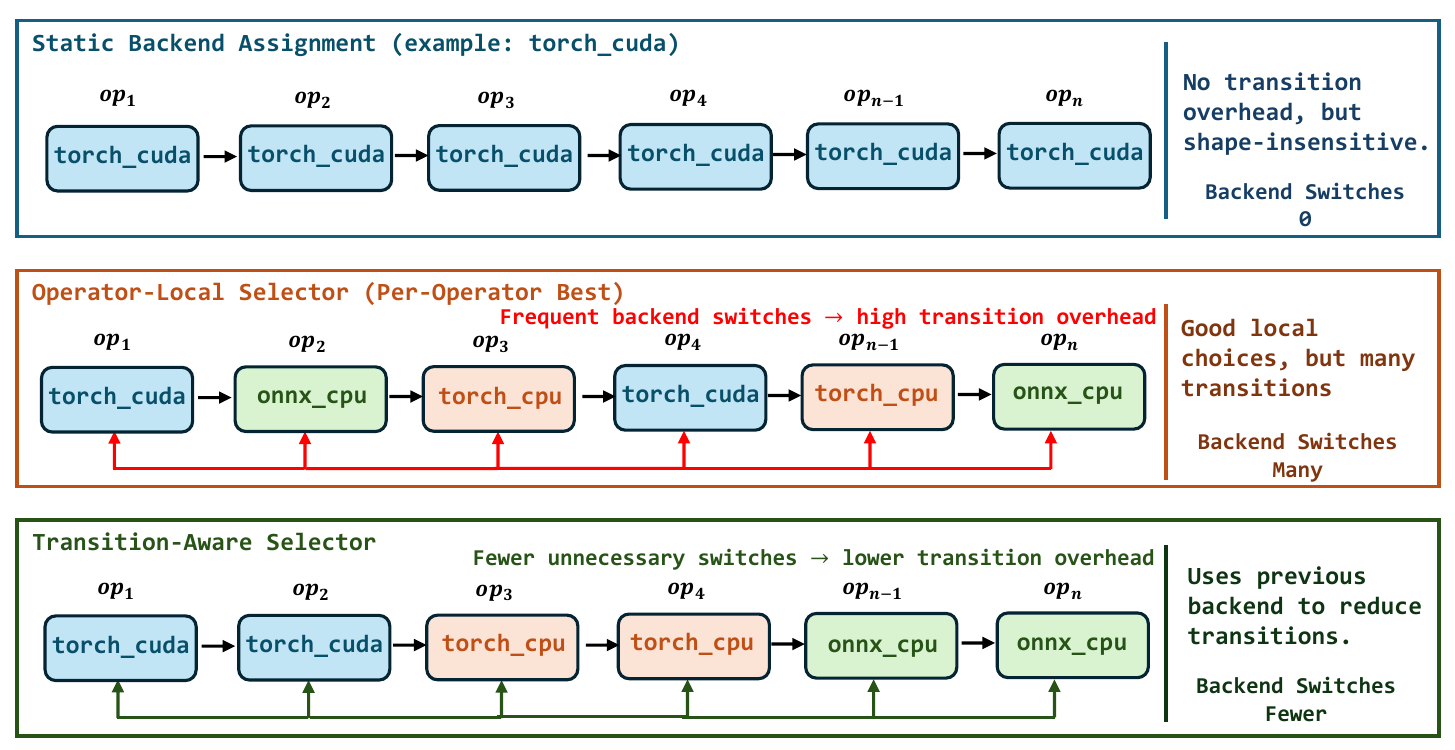}
    \caption{Comparison of static backend assignment, operator-local selection, and transition-aware dispatch for an ordered inference trace.}
    \label{fig:transition}
\end{figure}

The proposed methodology consists of model-trace collection, supported-operator extraction, exact-shape grouping, multi-backend benchmarking on Jetson, selector training, and measurement-backed trace replay. Fig.~\ref{fig:1stfigure} summarizes this pipeline, while Fig.~\ref{fig:transition} illustrates the three evaluated dispatch policies.

\subsection{Full-Model Inference Trace Collection}

Ordered operator traces are collected from full-model inference runs on Jetson. Each trace preserves the operator sequence together with the context and tensor-shape metadata required for backend selection. The traced models are Qwen2.5-0.5B~\cite{qwen25}, Qwen3-0.6B~\cite{qwen3}, TinyLlama-1.1B~\cite{tinyllama}, SmolLM2-135M and SmolLM2-360M~\cite{smollm2}, OLMo2-1B~\cite{olmo2}, and ModernBERT-base~\cite{modernbert}.

Trace collection uses synthetic tokenizer inputs with batch size 1 and FP32 data. Prefill sequence lengths are 16, 32, 64, and 128 tokens. Each input is constructed by repeating ``hello'' to the target length and tokenizing with maximum-length padding. For causal models, the decode trace contains one length-1 forward call rather than a complete autoregressive KV-cache generation loop.

Synthetic inputs are used to control sequence length and operator shape rather than to evaluate language-model accuracy. The collected metadata distinguishes prefill and decode phases, allowing phase-dependent shapes to be preserved during replay. The ordered trace files and extracted feature fields are included in the released artifact.

The dispatch scope covers Linear, multilayer perceptron (MLP) activation, root mean square normalization (RMSNorm), and layer normalization (LayerNorm). These classes contribute 9,584 ordered instances. Extracted features include model name, inference phase, batch size, sequence length, hidden and intermediate sizes, matrix dimensions $M$, $N$, and $K$, data type, and output size. Operators outside this scope retain a static assignment and are not selected dynamically.

The full model execution is used only to establish the model-derived order and operator contexts. Supported operators are subsequently evaluated through independent exact-shape measurements rather than by replacing operators inside an integrated model runtime.

\subsection{Exact-Shape Grouping}

Repeated instances are grouped by exact shape and execution context to avoid redundant benchmarking. Two instances share a group only when their operator type, phase, tensor dimensions, data type, and output size match. This produces 278 unique shape groups from the 9,584 supported instances.

Let $o_i$ denote the $i$-th supported operator and $x_i$ its feature vector. Exact-shape grouping maps $x_i$ to shape identifier $s_i$. Each shape-backend cost is measured once and reused whenever that shape appears during replay.

\subsection{Jetson Multi-Backend Benchmarking}

Each shape group is independently benchmarked using PyTorch eager CPU (\texttt{torch\_eager\_cpu}), PyTorch eager CUDA (\texttt{torch\_eager\_cuda}), and ONNX Runtime CPU (\texttt{onnxruntime\_cpu}). For each shape-backend pair, latency and active power are measured directly. An idle baseline is collected using the same repeat count and measurement procedure. Baseline-corrected power is
\[
P_i^{\mathrm{net}}(b)
=
P_i^{\mathrm{active}}(b)
-
P_i^{\mathrm{idle}}(b).
\]
Energy and per-operator energy-delay product (EDP) are
\[
E_i(b)
=
P_i^{\mathrm{net}}(b)
\times
\frac{L_i(b)}{1000},
\qquad
EDP_i(b)
=
E_i(b)
\times
\frac{L_i(b)}{1000},
\]
where latency $L_i(b)$ is measured in milliseconds. The resulting table contains 834 rows, corresponding to 278 shapes evaluated across three backends. All retained rows pass correctness validation and contain valid latency, power, energy, and EDP measurements.

\subsection{Backend Selector Training}

Selectors are trained separately for latency, energy, and EDP. The operator-local selector uses the current operator features, and its target is the backend with the lowest measured cost for the selected objective. The transition-aware selector additionally receives previous backend $b_{i-1}$ and evaluates candidate backend $b_i$ using
\[
C_i(b_i)
=
C_i^{op}(b_i)
+
C_i^{trans}(b_{i-1},b_i),
\]
where $C_i^{op}(b_i)$ is measured operator cost and $C_i^{trans}(b_{i-1},b_i)$ is the transition penalty estimated from directed backend-pair measurements. The transition term is zero when the backend is unchanged.

For the EDP objective, candidate cost is represented by an EDP proxy formed from the combined operator and transition latency and energy.

Directed transition measurements execute an operator on a source backend and then switch to a target backend. The resulting pair measurements estimate the switch latency and energy assigned to each directed candidate. For cost construction, every shape is expanded across three previous-backend states and three candidate backends, producing 2,502 candidate-cost rows. These include 834 same-backend candidates with zero transition cost and 1,668 directed backend-change candidates. The lowest-cost candidate for each shape and previous-backend state defines the label, yielding 834 transition-aware decision contexts.

The evaluated classifiers are Extra Trees and random forest with 200 estimators and a minimum leaf size of one, balanced logistic regression with 2,000 maximum iterations, and histogram-based gradient boosting (HistGBM). All use random seed 42. Model selection uses GroupKFold grouped by shape identifier, preventing candidate rows derived from the same exact shape from appearing in both training and validation folds. Models are selected by validation regret for each objective. HistGBM is selected for the reported transition-aware latency, energy, and EDP results.

\subsection{Transition-Aware Trace Replay}

The trained policies are evaluated through analytical replay of the model-derived operator traces. Replay preserves operator order but retrieves execution costs from the Jetson measurement tables rather than executing the connected model through an integrated mixed-backend runtime. Operators outside the dispatch scope retain a static assignment. For supported operators, static assignment uses one backend throughout the trace, operator-local selection predicts independently, and transition-aware selection uses the current features and previous backend. The latter policy is causal and greedy rather than globally sequence-optimal.

At each supported trace position, the policy predicts a backend, the corresponding measured operator cost is retrieved, and any directed transition penalty is added. The selected backend then becomes the previous-backend state for the next supported decision. Static assignments for unsupported operators are common across policies and are not modified by the selector. This design isolates the contribution of selective dispatch while retaining model-derived ordering.

For trace $\mathcal{T}=(o_1,o_2,\ldots,o_T)$ and backend sequence $(b_1,b_2,\ldots,b_T)$, replayed latency and energy are
\[
L_{\mathcal{T}}
=
\sum_{i=1}^{T}
\left[
L_i(b_i)
+
L_i^{trans}(b_{i-1},b_i)
\right],
\]
\[
E_{\mathcal{T}}
=
\sum_{i=1}^{T}
\left[
E_i(b_i)
+
E_i^{trans}(b_{i-1},b_i)
\right].
\]
Trace-level EDP is computed after accumulation as
\[
EDP_{\mathcal{T}}
=
E_{\mathcal{T}}
\times
\frac{L_{\mathcal{T}}}{1000}.
\]
This differs from the per-operator EDP used for label construction. Replay also records backend, device, and framework switches. Integrated dispatcher and model-level framework overheads are outside the replay scope.

\subsection{Experimental Setup}

Measurements use an NVIDIA Jetson Orin Nano running JetPack~6.2 and Linux for Tegra (L4T)~R36.4.3, with CUDA~12.6, PyTorch~2.8.0, ONNX Runtime~1.23.2, and Python~3.10.12. The device operates in NVIDIA 25~W mode without forced clocks, reflecting default clock behavior under the selected power mode.

Each exact-shape run uses three warmup iterations followed by 500 nominal measured iterations, with adaptive repetition for short operators. Retained rows contain between 10 and 27,077 iterations, with a median of 778. Power is sampled from the \texttt{VDD\_IN} rail using \texttt{tegrastats} at 50~ms intervals. Short operators are repeated over an extended window to obtain sufficient active and idle samples. The benchmark records mean, median, percentiles, range, and standard deviation as within-run statistics. Latency summaries include the mean, median, p50, p90, p95, p99, minimum, maximum, and standard deviation. Correctness is checked against PyTorch CPU reference output. FP32 tolerances are $\mathrm{rtol}=10^{-4}$ and $\mathrm{atol}=10^{-5}$ for RMSNorm, MLP activation, and LayerNorm, while Linear uses $\mathrm{atol}=10^{-3}$. Correctness status and absolute, mean, and relative error statistics are retained with each result.

\subsection{Evaluation Protocol}

The main evaluation compares static, operator-local, and transition-aware policies across seven model-derived traces. Gains are calculated relative to the best static policy for each objective. The main evaluation measures represented shape contexts. Leave-one-model-out evaluation uses LeaveOneGroupOut by model name, holding out one model during training and using it only for replay to evaluate model-level transfer.

\section{Evaluation Results}

\begin{figure}[t]
    \centering
    \includegraphics[width=\columnwidth]{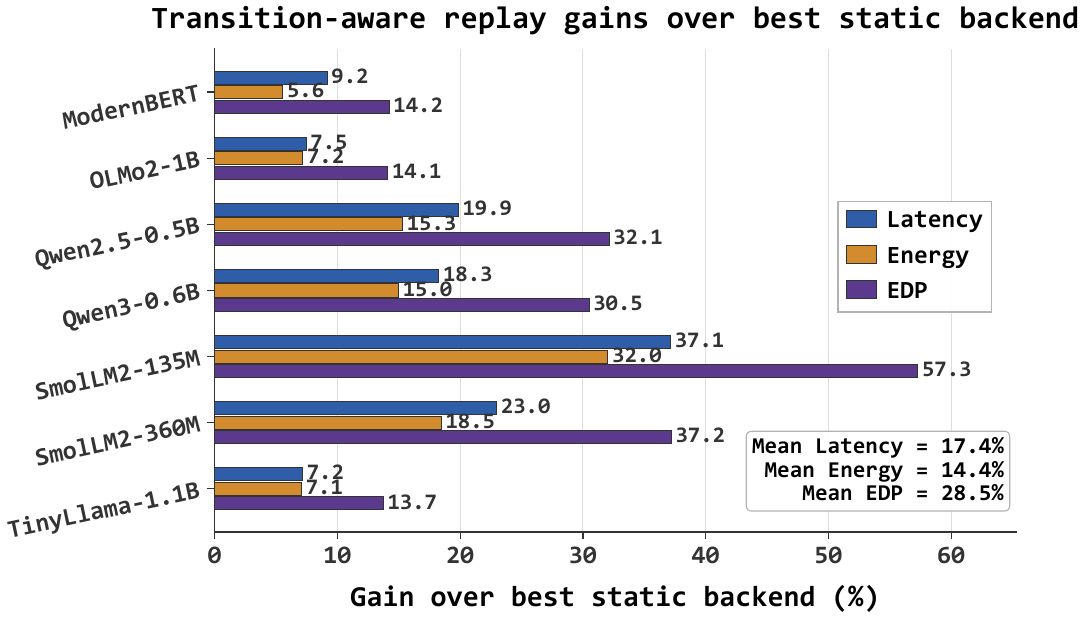}
    \caption{Transition-aware replay gains over the best static backend across seven model-derived operator traces.}
    \label{fig:transition_aware_gain}
\end{figure}

This section evaluates selector quality, measured-shape replay gains, the effect of transition awareness, and leave-one-model-out generalization. Table~\ref{tab:selector_accuracy_regret} reports accuracy and regret, Fig.~\ref{fig:transition_aware_gain} presents gains over static assignment, Fig.~\ref{fig:ablation} compares transition-aware and operator-local policies, and Table~\ref{tab:lomo_generalization} reports held-out-model results.

\begin{table}[t]
\centering
\caption{Selector accuracy and regret under grouped evaluation. Counts correspond to backend-selection decision contexts. Best accuracy values for each shared objective and lowest regret values within each objective are bolded.}
\label{tab:selector_accuracy_regret}
\footnotesize
\setlength{\tabcolsep}{4pt}
\renewcommand{\arraystretch}{1.08}
\begin{tabular}{ll l c c c}
\toprule
\textbf{Selector} & \textbf{Objective} & \textbf{Best model}
& \textbf{Contexts} & \textbf{Acc.} & \textbf{Regret (\%)} \\
\midrule
\multirow{3}{*}{Operator-local}
& Latency & RF & 278 & 0.8741 & \best{0.3407} \\
& Energy  & RF & 278 & 0.8705 & \best{0.1791} \\
& EDP     & RF & 278 & 0.8597 & \best{0.0052} \\
\midrule
\multirow{3}{*}{Transition-aware}
& Latency & HistGBM & 834 & \best{0.8921} & 0.4094 \\
& Energy  & HistGBM & 834 & \best{0.8897} & 0.1988 \\
& EDP     & HistGBM & 834 & \best{0.9005} & 0.0100 \\
\bottomrule
\end{tabular}

\begin{flushleft}
\scriptsize
\textit{RF: Random Forest}
\end{flushleft}
\end{table}

\subsection{Selector Accuracy and Regret}

Accuracy measures how often a selector predicts the lowest-cost backend for its decision context. Regret measures the relative cost gap between the selected backend and the lowest-cost candidate for that context; lower regret therefore indicates that incorrect predictions have limited objective-level impact.

As shown in Table~\ref{tab:selector_accuracy_regret}, the operator-local selector achieves accuracies of 0.8741, 0.8705, and 0.8597 for latency, energy, and EDP, respectively. Its regret remains below 0.35\% for every objective and reaches 0.0052\% for EDP. The transition-aware selector achieves accuracies of 0.8921, 0.8897, and 0.9005 over the expanded previous-backend decision contexts. Its regret is slightly higher than that of the operator-local selector but remains below 0.41\% for all objectives. Because the two selectors use different decision contexts and target costs, their accuracy values are not treated as a direct head-to-head improvement. Instead, the results show that both formulations maintain strong prediction quality, while transition awareness enables decisions conditioned on backend-switching cost.

\subsection{Main Trace-Replay Gains}

Figure~\ref{fig:transition_aware_gain} reports transition-aware replay gains relative to the best static backend policy for each model and objective. Positive gains are obtained across all seven model-derived traces for latency, energy, and EDP. The best static policies require 4,782.0~ms and 39.10~J per trace on average, while transition-aware replay reduces these values to 4,012.4~ms and 34.07~J, respectively. When per-trace gains are averaged, latency decreases by 17.4\%, energy by 14.4\%, and trace-level EDP by 28.5\%.

Gains are computed separately for each trace before averaging, preventing larger traces from dominating the reported mean percentages.

The magnitude of improvement varies with trace structure and exact-shape distribution. ModernBERT-base and OLMo2-1B produce EDP gains of 14.2\% and 14.1\%, respectively. Qwen2.5-0.5B and Qwen3-0.6B produce larger EDP gains of 32.1\% and 30.5\%. The strongest result occurs for SmolLM2-135M, where replayed latency, energy, and EDP improve by 37.1\%, 32.0\%, and 57.3\%, respectively. SmolLM2-360M also obtains an EDP gain of 37.2\%.

These results indicate that static assignment misses measurable opportunities when the preferred backend varies across supported operator shapes. The reported values quantify the potential of selective dispatch under the measured-cost replay model; they do not represent end-to-end measurements from an integrated mixed-backend runtime.

\subsection{Effect of Transition Awareness}

\begin{figure}[t]
    \centering
    \includegraphics[width=\columnwidth]{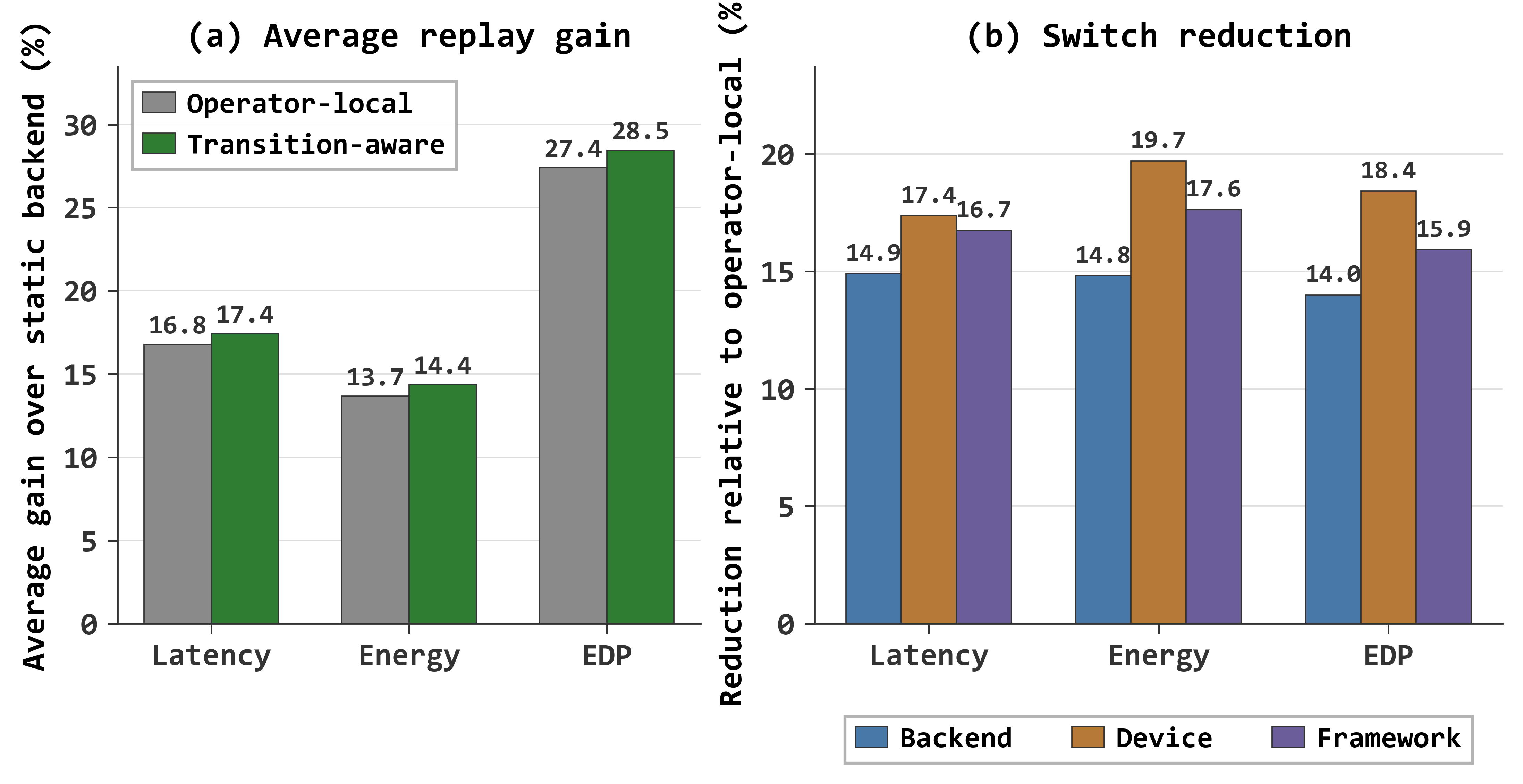}
\caption{Ablation of transition-aware dispatch against operator-local selection, showing average replay gain and switch reduction.}
    \label{fig:ablation}
    
\end{figure}

Figure~\ref{fig:ablation} compares transition-aware dispatch with operator-local selection. Operator-local selection already improves substantially over static assignment by adapting to individual operator shapes, but it does not account for the backend used by the preceding supported operator.

Transition awareness provides modest additional replay-cost gains. Average latency gain increases from 16.8\% to 17.4\%, energy gain from 13.7\% to 14.4\%, and EDP gain from 27.4\% to 28.5\%. Its primary effect is a reduction in switching. Relative to operator-local selection, transition-aware dispatch reduces backend switches by 14.9\%, 14.8\%, and 14.0\% for the latency, energy, and EDP objectives. Device switches decrease by up to 19.7\%, and framework switches decrease by up to 17.6\%.

The ablation therefore separates the benefit of shape-dependent selection from the additional benefit of previous-backend context. Most replay-cost improvement over static assignment comes from operator-level backend specialization, while transition awareness primarily produces a more stable backend sequence and avoids changes with insufficient local benefit.

\subsection{Leave-One-Model-Out Generalization}

Table~\ref{tab:lomo_generalization} evaluates model-level transfer using leave-one-model-out replay. In each run, one model is excluded from selector training and used only for evaluation. This setting tests whether learned shape and backend relationships transfer beyond the models represented during training.

Energy improves for all seven held-out models, while latency and EDP improve for six. The strongest held-out result occurs for SmolLM2-135M, with latency, energy, and EDP gains of 31.66\%, 26.97\%, and 49.75\%, respectively. Qwen2.5-0.5B, Qwen3-0.6B, SmolLM2-360M, OLMo2-1B, and TinyLlama-1.1B also improve across all three objectives. These results show consistent transfer across the six held-out decoder-oriented traces.

ModernBERT-base is the principal stress case. Its energy improves by 2.61\%, whereas latency and EDP gains are $-15.49\%$ and $-6.10\%$. This result suggests that transfer weakens when an encoder-style trace has an operator-shape distribution that differs from the decoder-oriented training traces. The leave-one-model-out evaluation therefore identifies both the transferability and the current boundary of the learned policy: related decoder-style traces generalize consistently, while structurally different workloads may require broader training coverage.

\begin{table}[t]
\centering
\caption{Leave-one-model-out trace-replay generalization. Gains are measured relative to the best static backend for each held-out model. Positive values indicate improvement.}
\label{tab:lomo_generalization}
\footnotesize
\setlength{\tabcolsep}{4pt}
\renewcommand{\arraystretch}{1.08}
\begin{tabular}{l r r r c}
\toprule
\textbf{Held-out model}
& \textbf{Latency}
& \textbf{Energy}
& \textbf{EDP}
& \textbf{All improved} \\
& \textbf{gain (\%)}
& \textbf{gain (\%)}
& \textbf{gain (\%)}
& \\
\midrule
ModernBERT-base  & \textcolor{red!70!black}{-15.49} & 2.61  & \textcolor{red!70!black}{-6.10} & \xmark \\
OLMo2-1B         & 7.38  & 7.15  & 14.01 & \cmark \\
Qwen2.5-0.5B     & 17.57 & 14.36 & 30.53 & \cmark \\
Qwen3-0.6B       & 17.95 & 14.09 & 28.79 & \cmark \\
SmolLM2-135M     & \best{31.66} & \best{26.97} & \best{49.75} & \cmark \\
SmolLM2-360M     & 22.30 & 18.43 & 36.82 & \cmark \\
TinyLlama-1.1B   & 6.20  & 7.09  & 11.94 & \cmark \\
\midrule
\textbf{Models improved}
& \best{6/7} & \best{7/7} & \best{6/7} & 6/7 \\
\textbf{Mean gain}
& 12.51 & 12.96 & \best{23.68} & -- \\
\bottomrule
\end{tabular}
\end{table}

\section{Future Work}

Future work will integrate the selector into a mixed-backend runtime, enabling direct validation of measurement-backed replay predictions during connected model execution. The benchmark will also be extended to additional transformer components, including attention, KV-cache operations, quantized operators, and fused kernels, together with optimized backends such as ONNX Runtime CUDA and TensorRT. These extensions will evaluate transition-aware dispatch across a broader range of deployment configurations while retaining the exact-shape measurement methodology.

\section{Conclusion}

This paper presented a measurement-backed trace-replay approach to transition-aware backend dispatch for edge transformer inference. The approach uses exact-shape Jetson measurements and previous-backend state to model selective eager-mode dispatch across PyTorch CPU, PyTorch CUDA, and ONNX Runtime CPU. Supported operators are selected dynamically, while operators outside the dispatch scope retain a static assignment. Across 9,584 supported operator instances and 278 exact shape groups from seven model-derived traces, transition-aware dispatch reduced replayed latency, energy, and EDP relative to the best static policy by 17.4\%, 14.4\%, and 28.5\% on average. Compared with operator-local selection, previous-backend context provided modest additional replay-cost gains while consistently reducing backend, device, and framework switches. Leave-one-model-out evaluation improved energy for all seven held-out models and improved latency and EDP for six. Overall, the results show that exact-shape measurements and backend-transition context provide useful signals for selective dispatch on heterogeneous edge platforms.

\bibliographystyle{IEEEtran}
\bibliography{related_work_edge_llm_dispatch}

\end{document}